\documentclass[conference]{IEEEtran}

\IEEEoverridecommandlockouts

\usepackage{xurl}
\usepackage{graphicx}
\usepackage{subcaption}
\usepackage{times,amsmath,amsfonts}
\usepackage{amsmath,amssymb,amstext}
\usepackage{multirow}
\usepackage{hhline}
\usepackage{multirow}
\usepackage{cite}
\usepackage{epstopdf}
\usepackage{bm}
\usepackage{algorithm} 
\usepackage{algorithmicx}
\usepackage{algpseudocode}
\usepackage{color}
\usepackage[lofdepth,lotdepth]{subfig}
\usepackage{soul}
\usepackage{float}
\usepackage{optidef}
\usepackage{adjustbox}

\begin{document}
\title{ML-Based Real-Time Downlink Performance Prediction in Standalone 5G NR Using Smartphones}

\author{Md Mahfuzur Rahman~\IEEEmembership{Senior Member,~IEEE}, Jareen Shuva, Nishith Tripathi ~\IEEEmembership{Senior Member,~IEEE},\\ Jeffrey H. Reed~\IEEEmembership{Fellow,~IEEE} , and Lingjia Liu ~\IEEEmembership{Fellow,~IEEE}
\thanks{M. M. Rahman, J. Shuva, N. Tripathi, J. H. Reed, and L. Liu are with Wireless@Virginia Tech, Bradley Department of Electrical and Computer Engineering, Virginia Tech, VA, USA. Corresponding author's e-mail: ljliu@vt.edu.}}
%

\maketitle

\begin{abstract}

We propose a machine learning (ML)-based framework for downlink performance prediction in 5G networks using real-time measurements from commercial off-the-shelf (COTS) user equipment (UE). Our experimental platform integrates the srsRAN 5G New Radio (NR) stack deployed on a Dell desktop serving as the 5G next generation nodeB (gNB), operating at 3.4 GHz. Two Google Pixel 7a smartphones are used to collect physical layer characteristics such as channel quality indicator (CQI), modulation and coding scheme (MCS), bit rate, transmission time interval (TTI), and block error rate (BLER), which are leveraged as predictors in model training. We use commercial-grade traffic generation tools, including Ookla, for stationary and mobility measurements under line-of-sight (LOS) and non-line-of-sight (nLOS) conditions. Test data includes global Ookla servers (e.g., USA, Portugal, Ghana, Egypt, Japan), iperf TCP/UDP data, and video streaming sessions from YouTube. To analyze inter-user interference, we also include scenarios with multiple UEs at the same location. We evaluate the predictive performance of five supervised regression models — linear regression, decision tree regression, random forest regression, extreme gradient boosting (XGBoost), light gradient boosting machine (LightGBM). Our results demonstrate that throughput and BLER can be accurately predicted
using COTS hardware and standard ML techniques in diverse real-world 5G scenarios.

\end{abstract}


\section{Background and Related Research}
\subsection{Introduction}

In recent years, the cellular landscape has been significantly transformed by the emergence of the Open Radio Access Network (ORAN) paradigm \cite{tripathi2025fundamentals}. ORAN promotes openness, software-driven architecture, network programmability, interoperability, and embedded intelligence—principles that have matured in wired networks through Software-Defined Networking (SDN) but are still evolving in mobile communications. Translating the Open RAN vision into practical, intelligent, and performance-optimized wireless systems presents several technical challenges, including: (i) acquiring high-quality datasets for training AI/ML models, (ii) establishing non-intrusive and reproducible test environments for model evaluation, (iii) ensuring automated and continuous software validation, and (iv) achieving reliable integration across diverse network components and platforms \cite{Colosseum}.

To address these challenges, open-source software platforms such as the Software Radio Systems Radio Access Network (srsRAN) and OpenAirInterface (OAI) have emerged as widely adopted tools for 5G New Radio (NR) experimentation. These platforms enable full-stack implementations that are compliant with 3GPP specifications and compatible with Commercial Off-The-Shelf (COTS) User Equipment (UE) and Software Defined Radios (SDRs) \cite{srsRAN4G, openairinterface}. Their modular architectures and low-cost deployment options allow researchers to prototype, configure, and evaluate 5G networks under realistic, controllable conditions. In particular, srsRAN has become a preferred testbed for many researchers due to its robust support for experimentation, reproducibility, and ease of integration with ML pipelines.

With increasing demands for high-throughput and low-latency services, predicting 5G network performance has become essential for adaptive and efficient radio resource management. 5G throughput is influenced by a complex combination of physical-layer metrics, including Reference Signal Received Power (RSRP), Signal-to-Interference-plus-Noise Ratio (SINR), and Modulation and Coding Scheme (MCS) levels \cite{teixeira2021predictive, perveen2022dynamic}. To address the challenge of modeling these nonlinear relationships, ML techniques have been extensively applied to forecast network throughput using real-time radio measurements \cite{rehmani2023machine, eyceyurt2022machine, elsherbiny2020throughput}. Such models enhance network intelligence by providing predictive insights that support scheduling, handover, and link adaptation decisions based on current channel conditions.

srsRAN-based testbeds have played a pivotal role in the development and validation of these ML models. Their ability to provide fine-grained control over experimental variables and support reproducible scenarios makes them ideal platforms for performance prediction studies \cite{Ye, alves2024experimental}. For instance, Random Forest and k-Nearest Neighbor algorithms have demonstrated high predictive accuracy, achieving $R^2$ scores above 0.9 for both uplink and downlink throughput in srsRAN-based LTE environments \cite{eyceyurt2022machine, elsherbiny2020throughput}. In addition to throughput prediction, srsRAN has been extended to support advanced research in protocol fuzzing \cite{mishra2023scaling}, security evaluation \cite{Shen}, and diagnostic analytics \cite{forbes2023closer}.

Several studies have contributed significantly to this field. Rehmani et al. \cite{rehmani2023machine} validated the use of Random Forest and regression models on real-world LTE data, confirming that ML techniques offer reliable tools for adaptive network management. Dias et al. \cite{dias2022rsrp2} developed an srsRAN-based testbed to predict RSRP and assist with LTE planning and link budget analysis. Similarly, Koenig et al. \cite{koenig2021throughput} created a cross-technology throughput prediction pipeline applicable to both LTE and 5G systems. Sinha et al. \cite{sinha2022prototyping} introduced a flexible software-defined radio testbed built on srsRAN and OAI to support experimentation across 4G and 5G systems in Open RAN environments.

Machine learning models used for throughput prediction in 5G include Linear Regression, Decision Tree Regression, Random Forest Regression, XGBoost, and Light GBM. Linear Regression has been used in multiple studies to model throughput from basic link quality indicators. For example, Elsherbiny et al. \cite{elsherbiny20204g} applied Linear Regression to COTS UE measurement data and showed that even simple statistical models can provide real-time throughput estimation. Decision Tree Regression, capable of capturing nonlinear feature interactions, has been applied to predict inter-frequency measurements, which are crucial for efficient handovers and planning \cite{sonnert2018predicting}. Its interpretability makes it well-suited for scenarios where transparency of the decision process is important. Random Forest Regression, as an ensemble method, has consistently outperformed classical models in LTE tasks. Chang and Baliga \cite{chang2021development} showed that Random Forest Regression improves path loss prediction accuracy, especially in dense urban deployments. Additionally, Random Forest Regression has been employed to predict RSRP and generate detailed coverage maps using data collected from smartphones [4].

The use of COTS UEs, such as modern smartphones, underscores the practical value of ML-based models in real-world 5G deployments. These devices allow extensive data collection without requiring expensive measurement equipment. When combined with open-source software platforms like srsRAN, they enable highly customizable testbeds that bridge the gap between theoretical research and field implementation. This study builds upon these prior efforts by developing and evaluating a real-time LTE testbed using srsRAN and COTS smartphones. Our focus is not on enhancing physical-layer throughput and BLER directly, but rather on accurately predicting downlink throughput and BLER based on observable radio conditions. We implement a lightweight ML pipeline using models such as Linear Regression, Decision Tree Regression, Random Forest Regression, XGBoost, and Light GBM, trained on live measurements that include channel quality indicator (CQI), MCS, Bit rate, transmission time interval (TTI), and block error rate (BLER). The goal is to enable intelligent estimation of throughput and BLER without active probing, which can support proactive scheduling, link adaptation, and quality-of-service management in next-generation networks. Importantly, while these models provide strong predictive capabilities, they do not modify or improve the underlying physical performance of the system.
\subsection{Related Research}

Minovski et al. \cite{Minovski} proposed a machine learning-based approach to predict LTE and 5G throughput using RF features such as RSRP, RSRQ, and SINR. They conducted passive and active measurements over commercial operator networks across diverse real-world scenarios, including urban, rural, and underground environments. Data collection was done using TEMS Pocket and a custom TWAMP-based probing tool, enabling accurate throughput labeling without the use of a dedicated testbed. Raca et al. \cite{Raca} developed an ML/DL-based framework for throughput prediction in LTE networks using device- and network-level metrics. They evaluated Random Forest, SVM, and LSTM models, showing that deep learning—especially LSTM—offers superior accuracy under highly dynamic conditions. Their data was collected from commercial devices using Android APIs and specialized tools, and they demonstrated improved QoE in video streaming with prediction-assisted adaptation. Al-Thaedan et al. \cite{Al-Thaedan} investigated downlink throughput prediction in 4G-LTE networks using real-world measurement data collected from three commercial operators in an urban environment via drive testing. They evaluated multiple machine learning models, including Support Vector Regression (SVR), Linear Regression, K-Nearest Neighbors (KNN), and Decision Tree Regression, using features such as RSRP, RSRQ, SINR, and GPS coordinates. Their results showed that KNN and DTR significantly outperformed SVR and LR, achieving up to 99\% R² in some scenarios. The study demonstrates the feasibility of using ML models for throughput forecasting but relies entirely on operator-driven environments and offline data post-processing. Basit et al. \cite{Basit} studied fine-grained throughput prediction using machine learning over commercial LTE and 5G networks, collecting data from a 1000 km drive with Android smartphones and Accuver tools. They evaluated models like Multi-Layer Perceptron (MLP) and Gradient Boosted Decision Trees (GBDT) but found poor accuracy on app-level metrics, attributing this to conflicting input samples and unpredictable network behavior at 100 ms resolution. Eyceyurt et al. \cite{Eyceyurt} proposed a machine learning-based framework to predict uplink throughput in LTE networks using physical layer features such as RSRP, RSRQ, and SNR. They collected drive-test data from multiple cities and evaluated five models—Linear Regression, Gradient Descent, Gradient Boosting, Decision Tree, and KNN—across various radio environments. Their results showed that Decision Tree and KNN provided the most accurate predictions, with R² values exceeding 0.9 in suburban settings. The study emphasizes that feature selection and location-specific variability significantly influence model performance, especially in metropolitan areas with complex interference patterns. Abdiel \cite{abdiel2022forecasting} proposed an LTE throughput forecasting framework to enhance both network efficiency and strategic business decision-making. A technical report from Clemson University \cite{clemson2022machine} compared various ML and deep learning models, demonstrating that deep neural networks more effectively capture nonlinearities in throughput behavior.

While aforementioned works have demonstrated the potential of machine learning for throughput prediction in LTE and 5G networks, these studies primarily rely on measurements from commercial operator networks and data collected through drive testing or passive logging. Such environments, while realistic, limit control over radio conditions and hinder the reproducibility of experimental results. In contrast, our study leverages a fully controlled 5G testbed built using srsRAN and commercial off-the-shelf (COTS) smartphones. This environment enables repeatable experiments, fine-grained manipulation of network parameters, and seamless integration of ML-based algorithms. Additionally, we focus on lightweight, interpretable models—Linear Regression, Decision Tree, Random Forest, XGBoost, and LGBM — that offer a favorable trade-off between accuracy and computational efficiency. Unlike deep models explored in prior work, our approach supports real-time, on-device inference, making it suitable for deployment in resource-constrained scenarios. By enabling algorithmic integration within a live testbed, our framework offers a reproducible and practical path for ML-driven throughput and BLER prediction and optimization under tunable 5G conditions.

\section{Motivation and Contribution}
\subsection{Motivation}
As mobile networks evolve toward greater flexibility, intelligence, and openness, the Open Radio Access Network (Open RAN) paradigm has emerged as a transformative framework. It emphasizes software-defined architecture, interoperability, and programmability—principles aimed at accelerating innovation and lowering the barrier to experimentation. However, translating this vision into practical deployments requires overcoming key challenges such as acquiring high-quality datasets, enabling non-intrusive experimentation, and validating AI/ML models in realistic environments. Accurate prediction of throughput and channel characteristics such as BLER in 5G networks is critical for proactive resource management, link adaptation, and quality-of-service provisioning. Physical-layer metrics such as TTI, CQI, BLER and MCS directly influence throughput, yet their complex and nonlinear interactions are not easily captured by traditional rule-based algorithms. Similarly, BLER also directly depends on same sets of physical layer metrics such as TTI, Bit rate, CQI, MCS.

ML offers a powerful alternative by modeling these relationships from real-world data. However, validating such models requires test environments that are both realistic and controllable. Commercial networks are unsuitable due to their operational constraints and limited observability. This creates a strong need for low-cost, reproducible 5G testbeds. Open-source platforms like srsRAN, combined with COTS smartphones and SDRs, provide a promising foundation for addressing this gap. These platforms enable end-to-end experimentation while offering fine-grained control over network parameters. Yet, despite their growing popularity, relatively few studies have systematically integrated COTS UEs with srsRAN to collect measurement data and evaluate ML-driven throughput prediction models under real-time conditions. This work is motivated by the need to bridge that gap: to demonstrate how open-source 5G SA stacks and readily available hardware can support practical, reproducible research into data-driven network intelligence—aligning both with the vision of Open RAN and the operational realities of 5G.

\subsection{Contribution}
To demonstrate the novelty and practical value of our approach, we summarize the main contributions of this work as follows:
\begin{itemize}
    \item \textbf{COTS-Based 5G SA Testbed Deployment:} We developed a realistic 5G SA experimental setup using srsRAN and a Dell desktop gNB, with a Google Pixel 7a acting as the COTS UE. This setup allows practical data collection and validation under controlled yet realistic wireless conditions at 3.4 GHz.
    
    \item  \textbf{Throughput Prediction Framework:} We implemented and compared five regression models—Linear Regression, Decision Tree Regression, Random Forest Regression, XGBoost, and LGBM—for predicting downlink throughput based on physical-layer measurements at the UE.

    \item  \textbf{BLER Prediction Framework:} We have utilized those five regression models for downlink BLER prediction based on physical-layer measurement.
    
    \item  \textbf{Feature Importance Analysis with Physical Insights:} We performed detailed feature importance analysis, identifying MCS as the most predictive feature for throughput analysis whereas TTI is the most predictive feature for BLER analysis.     
  
\end{itemize}

\section{Regression Model Formulations}

This section outlines the underlying mathematical formulations and characteristics of the three regression models used for 5G throughput and BLER prediction: Linear Regression, Decision Tree Regression, and Random Forest Regression. These models were chosen to balance prediction performance, computational efficiency, and interpretability across varying levels of model complexity.

\subsection{Linear Regression}

Linear Regression is a fundamental model that assumes a linear relationship between the input features and the output variable. It fits a weight vector $\mathbf{w}$ by minimizing the squared loss between the predicted and actual values over all training samples:

\begin{equation}
\min_{\mathbf{w}} \left( \sum_{i=1}^{n} (y_i - \mathbf{w}^\top \mathbf{x}_i)^2 \right)
\end{equation}

In this formulation, $\mathbf{x}_i$ represents the input feature vector for the $i$-th sample, $y_i$ is the true throughput, and $\hat{y}_i = \mathbf{w}^\top \mathbf{x}_i$ is the predicted throughput. The model parameters $\mathbf{w}$ are estimated via closed-form solutions or iterative methods such as gradient descent.

Linear Regression is computationally lightweight and easy to interpret, making it a suitable baseline for comparison. However, due to its linearity assumption, it may not capture the complex nonlinear dependencies that often arise in wireless networks due to phenomena like fading, interference, or dynamic bandwidth allocation.

\subsection{Decision Tree Regression}

Decision Tree Regression is a non-parametric model that recursively partitions the input space into disjoint regions based on feature thresholding. The goal is to split the dataset in a way that minimizes the variance of the target variable within each region. At each decision node, the algorithm chooses a feature $j$ and a threshold $t$ that result in the greatest reduction in prediction error:

\begin{equation}
\min_{j, t} \left( \sum_{x_i \in R_1(j, t)} (y_i - \bar{y}_{R_1})^2 + \sum_{x_i \in R_2(j, t)} (y_i - \bar{y}_{R_2})^2 \right)
\end{equation}

Here, $R_1$ and $R_2$ are the subsets of data partitioned by the condition $x_j \leq t$ and $x_j > t$, respectively. The predicted output in each region is the mean throughput $\bar{y}_{R_k}$ of samples in that region. 

Decision Trees can capture non-linear interactions between features and are highly interpretable. They are particularly useful when feature interactions are hierarchical or when thresholds (e.g., specific SNR or bandwidth levels) are meaningful. However, without constraints such as maximum depth or minimum samples per leaf, decision trees are prone to overfitting.

\subsection{Random Forest Regression}

Random Forest is an ensemble learning method that constructs a large number of decision trees, each trained on a random subset of the training data (via bootstrapping) and a random subset of features. The final prediction is the average of the outputs from all individual trees:

\begin{equation}
\hat{y} = \frac{1}{B} \sum_{b=1}^{B} T_b(\mathbf{x})
\end{equation}

In this equation, $T_b(\mathbf{x})$ denotes the output of the $b$-th decision tree, and $B$ is the total number of trees in the ensemble. This technique reduces the variance inherent in individual decision trees and typically results in improved generalization on unseen data.

Random Forests are well-suited for capturing nonlinear relationships and complex feature interactions common in 5G networks, such as the combined impact of modulation schemes and bandwidth allocation. They also provide internal measures of feature importance, which are useful for identifying dominant physical-layer parameters in wireless performance prediction.
\subsection{Extreme Gradient Boosting (XGBoost) Regression}

XGBoost is a scalable and efficient gradient boosting framework that builds an ensemble of regression trees sequentially. Each new tree aims to correct the residual errors of the previous trees by minimizing a regularized objective function that balances training loss with model complexity. The final prediction is the sum of the outputs from all trees:

\begin{equation}
\hat{y} = \sum_{b=1}^{B} f_b(\mathbf{x}), \quad f_b \in \mathcal{F}
\label{eq:xgboost}
\end{equation}

In this equation, $f_b(\mathbf{x})$ denotes the output of the $b$-th regression tree, and $\mathcal{F}$ represents the space of all possible trees. XGBoost employs second-order optimization, utilizing both first- and second-order derivatives of the loss function, which enhances convergence speed and predictive accuracy.

XGBoost is particularly suitable for modeling high-dimensional and structured data often observed in 5G networks. Its built-in regularization mechanisms and tree pruning techniques help prevent overfitting while capturing complex feature interactions—such as those involving modulation, coding, and scheduling. Additionally, XGBoost provides internal measures of feature importance, facilitating model interpretability in wireless performance prediction tasks.

\subsection{Light Gradient Boosting Machine (LightGBM) Regression}

LightGBM is a high-efficiency gradient boosting framework designed for speed and scalability. Unlike traditional level-wise tree growth, LightGBM employs a leaf-wise tree growth strategy, which expands the leaf with the maximum split gain. This approach leads to deeper trees and often better accuracy with fewer iterations. The model prediction is the sum of outputs from all trees:

\begin{equation}
\hat{y} = \sum_{b=1}^{B} f_b(\mathbf{x}), \quad f_b \in \mathcal{F}_{\text{leaf-wise}}
\label{eq:lgbm}
\end{equation}

Here, $f_b(\mathbf{x})$ is the prediction from the $b$-th leaf-wise regression tree, and $\mathcal{F}_{\text{leaf-wise}}$ denotes the space of leaf-wise trees. LightGBM utilizes histogram-based binning for continuous features, which reduces memory usage and accelerates training.

LightGBM is well-suited for 5G datasets with large sample sizes and high feature dimensionality. Its ability to handle categorical features natively and focus on informative splits makes it effective in capturing nonlinear relationships among physical-layer parameters. Like XGBoost, LightGBM offers feature importance metrics that are valuable for identifying key predictors in wireless link performance analysis.

In this work, we formulate 5G downlink throughput and BLER prediction as a supervised regression task. Each input sample is represented as a feature vector:

\begin{align*}
    &\hspace{2cm} \mathbf{x_{\text{throughput}}} = \left[\text{CQI}, \text{MCS}, \text{TTI}, \text{BLER}\right] ,\\
    and \\
    \vspace{10pt}
    &\hspace{2cm} \mathbf{x_{\text{BLER}}} = \left[\text{CQI}, \text{MCS}, \text{TTI}, \text{Bit rate}\right].
\end{align*}

The corresponding output is measured as the downlink throughput/BLER, y. The goal is to learn a function $\hat{y}=f(x)$ that minimizes the prediction error between actual and estimated values. The dataset was split into 80\% training and 20\% testing sets. All input features were standardized using z-score normalization to ensure consistent scale across dimensions and improve model convergence. Z-score normalization (also called standardization) is a method of rescaling data so that it has a mean of 0 and a standard deviation of 1. This conversion is very useful when features have different scales and improves the performance of many ML algorithms that are sensitive to scale. Without normalization, features with larger numerical ranges can dominate the learning process, potentially leading to suboptimal convergence behavior and model performance. By applying Z-score normalization, the learning algorithm can converge more efficiently and produce more reliable results across features of varying magnitudes. We have used StandardScaler library from the Scikit-learn package to implement this Z-score normalization. To assess model performance, we use the following standard regression metrics:

(a) Mean Squared Error (MSE)\\
MSE measures the average squared difference between actual and predicted throughput:
\begin{equation}
\text{MSE} = \frac{1}{n} \sum_{i=1}^{n} (y_i - \hat{y}_i)^2
\end{equation}
Its very sensitive to large errors due to squaring and it has a unit in $Mbps^2$. 

(b) Root Mean Squared Error (RMSE)\\
RMSE provides a more interpretable error metric (same unit as target):
\begin{equation}
\text{RMSE} = \sqrt{\text{MSE}} = \sqrt{\frac{1}{n} \sum_{i=1}^{n} (y_i - \hat{y}_i)^2}
\end{equation}
The RMSE quantifies the average magnitude of prediction errors in Mbps. Lower RMSE values correspond to more accurate predictions, as they indicate smaller deviations between predicted and actual values.

(c) $R^2$ Score\\
$R^2$ Score indicates the proportion of variance in throughput explained by the model:
\begin{equation}
R^2 = 1 - \frac{\sum_{i=1}^{n} (y_i - \hat{y}_i)^2}{\sum_{i=1}^{n} (y_i - \bar{y})^2}
\end{equation}
The coefficient of determination ($R^2$) measures the proportion of variance in the dependent variable that is predictable from the independent variables. A higher $R^2$ signifies improved model performance, with the ideal value of 1 indicating perfect prediction accuracy.


In our simulation five regression models were implemented using the scikit-learn library: (a) Linear regression: A baseline model assuming a linear relationship between inputs and throughput, (b) Decision tree regression: A non-parametric model that partitions the feature space and fits constant values in each region, and (c) Random forest regression: An ensemble of decision trees trained on bootstrapped data and random feature subsets, d) XGBoost, and e) LGBM -known for capturing nonlinear relationships and reducing overfitting. To interpret the model’s internal behavior, we extract feature importance scores from the LGBM model. These scores indicate how much each feature contributes to reducing prediction error across trees. This analysis is critical for understanding which radio parameters most affect throughput in practical 5G deployments.

\section{System Model and Data Collection}

The experimental testbed for evaluating the proposed downlink performance prediction framework was developed using a custom-built, standalone 5G NR system conforming to Open RAN principles. The RAN components were deployed using the open-source srsRAN 5G stack, which enables modular and flexible configuration of gNB and core functions. A Dell desktop powered by an Intel i7 CPU and 32 GB RAM hosted the 5G gNB and interfaced with the AMF component of a lightweight Open5GS core. The host machine ran Ubuntu 24.04 LTS with low-latency kernel patches to support timing-critical over-the-air operations. The system operated in the sub-6 GHz band, centered at 3.4 GHz, using an Ettus Research USRP B210 software-defined radio as the RF front end. The USRP was synchronized with the host via USB 3.0 and configured for 20 MHz channel bandwidth, with appropriate TX/RX antenna gain settings for reliable signal propagation in both indoor lab and outdoor environments. Two Google Pixel 7a smartphones running Android 14 served as 5G UEs for user-side measurements. The extracted features included CQI, MCS, TTI, bit rate, and BLER, which were used as input-output pairs for model training and testing.

\begin{figure}[htbp]
\centering
\includegraphics[width=3.5in, height=1.2in]{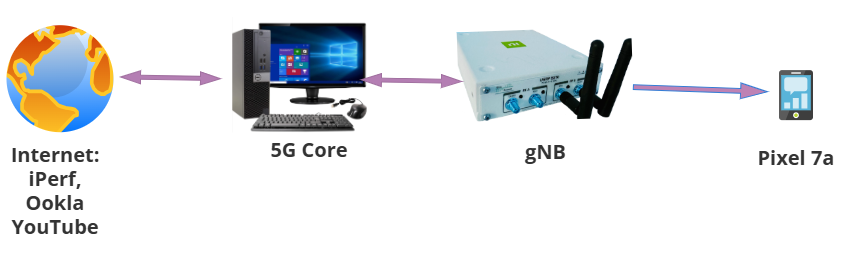}
\caption{Standalone 5G testbed using a commercial smartphone}
\label{system-model}
\end{figure}

\subsection{Data Collection}
We developed standalone 5G testbeds using both srsRAN 5G and OpenAirInterface (OAI) stacks in indoor and outdoor settings. However, all analysis in this work is restricted to the srsRAN-based testbed. As summarized in \ref{data}, data collection was conducted under various link conditions, including LOS, nLOS, stationary, and with mobility scenarios. We used commercial-grade applications such as Ookla Speedtest, Magic iPerf, and YouTube to generate downlink traffic.

\begin{table}[htbp]
    \centering
    \caption{Testing tools and scenarios}
    \begin{adjustbox}{width=0.45\textwidth}
    \begin{tabular}{|l|c|c|c|c|}
        \hline
        \textbf{Test Type} & \textbf{UE type} & \textbf{Scenario} & \textbf{Protocol}& \textbf{Testing Mode}\\
        \hline
        Ookla & Pixel 7a & Indoor \& Outdoor & TCP & Static \& Mobility\\
        \hline
        YouTube & Pixel 7a & Indoor \& Outdoor & UDP & Static \& Mobility\\
        \hline
        iPerf & Pixel 7a & Indoor \& Outdoor & TCP \& UDP & Static \& Mobility\\
        \hline           
    \end{tabular}
    \end{adjustbox}
    \label{data}
\end{table}

\subsubsection{Ookla Testing}
To install a dedicated Ookla server on a local lab PC, a public IP address is required so that the server appears in the Speedtest client application. Due to institutional security policies, public IP sharing is restricted and subject to administrative approval. While using a local server could improve measurement consistency, we relied on both local and cross-continent Ookla servers for our tests. Regional iPerf servers within a 50 km radius were also utilized. Global Speedtest servers included those located in North America, Ghana, Egypt, Saudi Arabia, Portugal, and Japan. 

\begin{figure*}[htbp]
    \centering
    \begin{subfigure}[b]{0.3\linewidth}
        \centering
        \includegraphics[width=2in, height=2in]{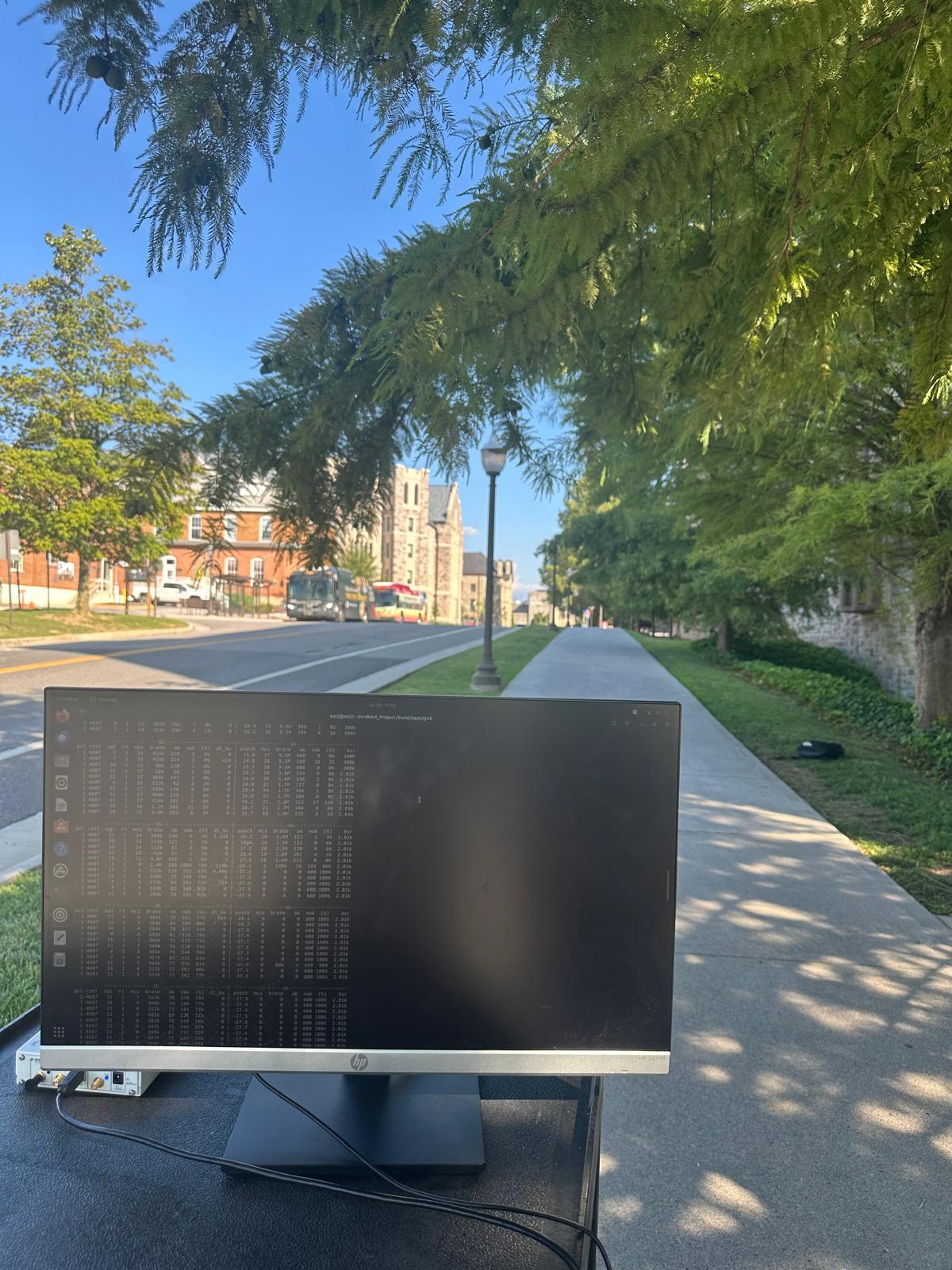}
        \caption{Outdoor testing}
        \label{ookla_outdoor}
    \end{subfigure}
    %
    %
    \begin{subfigure}[b]{0.3\linewidth}
        \centering
        \includegraphics[width=2in, height=2in]{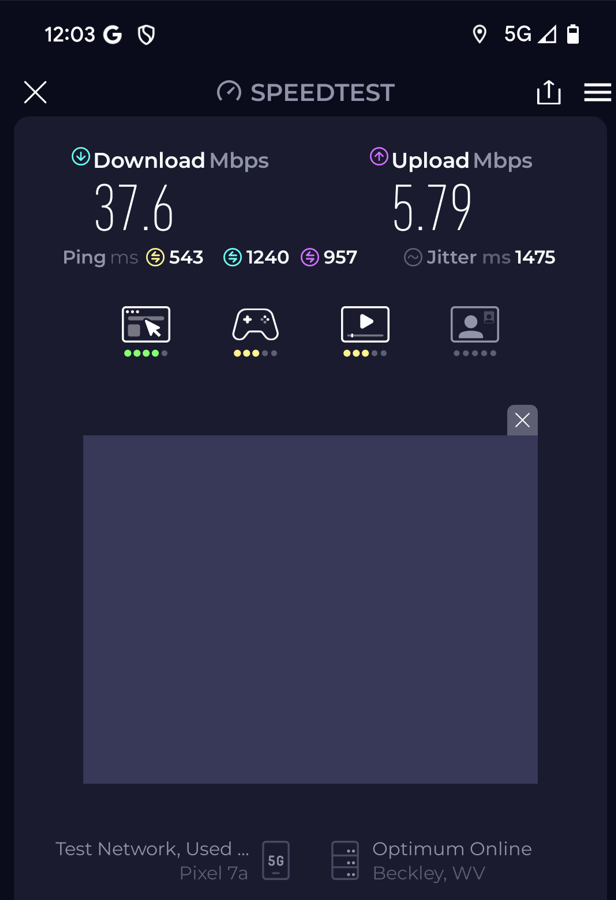}
        \caption{Local testing (Beckley, WV, USA)}
        \label{ookla_becky}
    \end{subfigure}
    %
    \begin{subfigure}[b]{0.3\linewidth}
        \centering
        \includegraphics[width=2in, height=2in]{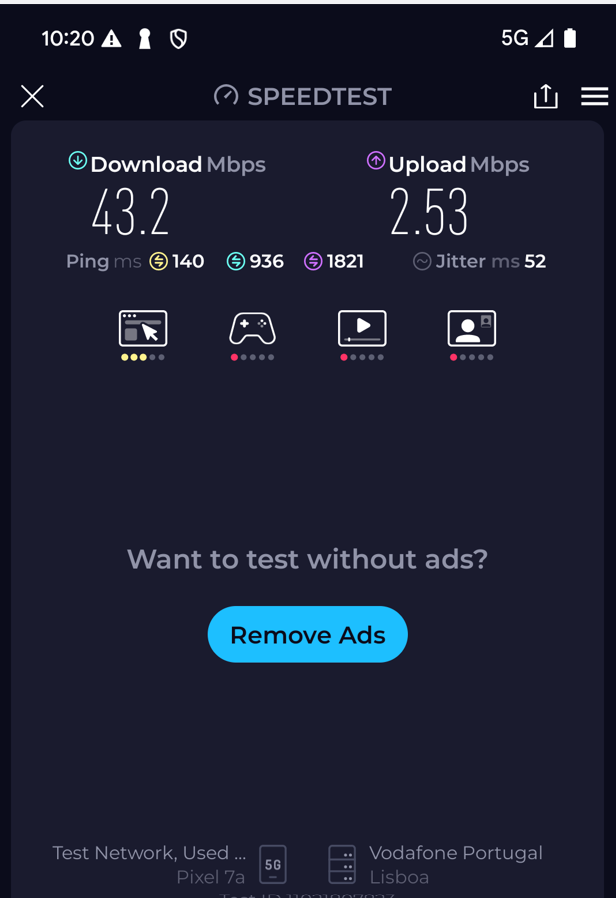}
        \caption{Cross-continent testing (Lisbon, Portugal)}
        \label{ookla_lisbon}
    \end{subfigure}
    \caption{Ookla speed testing using local and cross-continent servers}
    \label{ookla}
\end{figure*}

\subsubsection{iPerf Testing}
Both TCP and UDP protocols were used for iPerf-based downlink testing via the Magic iPerf app installed on the UE. Publicly available iPerf servers located in Ashburn, VA, and Atlanta, GA were selected to introduce backhaul diversity on iPerf DL testing. These tests allowed us to evaluate protocol-specific throughput patterns and sensitivity to backhaul network variability across different transport layers.

\begin{figure}[htbp]
    \centering
    \begin{subfigure}[b]{\linewidth}
        \centering
        \includegraphics[width=2.5in, height=2in]{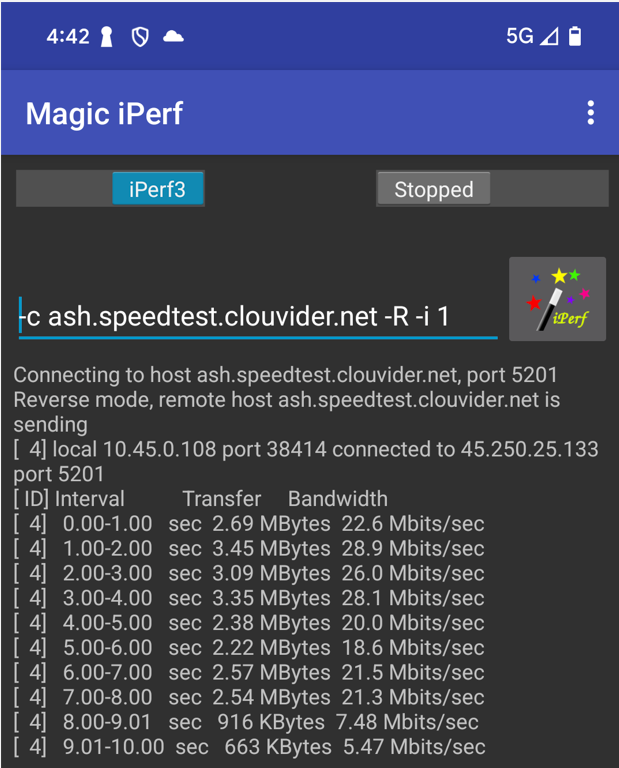}
        \caption{iPerf TCP testing}
        \label{iperf_tcp}
    \end{subfigure}
    
    \vspace{0.5cm}
    
    \begin{subfigure}[b]{\linewidth}
        \centering
        \includegraphics[width=2.5in, height=2in]{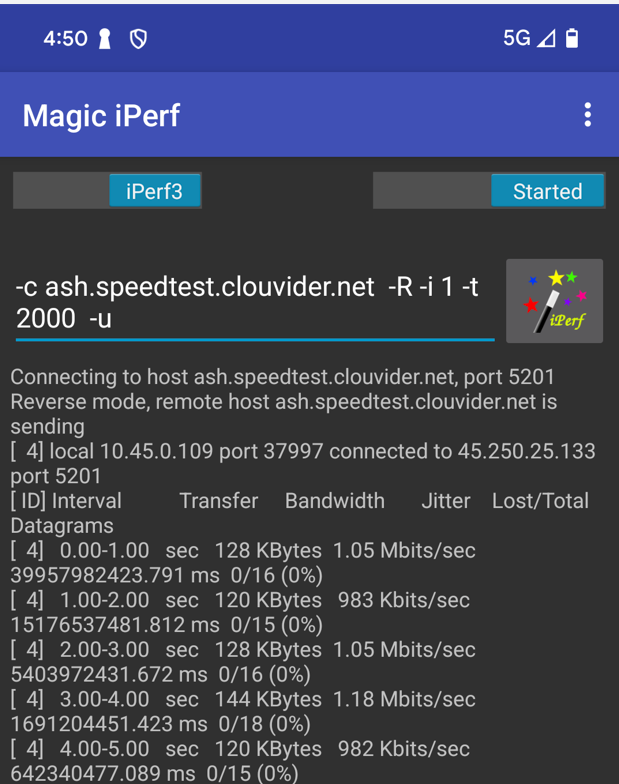}
        \caption{iPerf UDP testing}
        \label{iperf_udp}
    \end{subfigure}
    \caption{iPerf downlink performance evaluation}
    \label{iperf}
\end{figure}

\subsubsection{YouTube Testing}
To evaluate performance under application-driven streaming workloads, we used YouTube to generate downlink feature vectors in both indoor and outdoor conditions. The dataset includes scenarios involving YouTube Live, 4K, and 8K video playback. Unlike iPerf, YouTube traffic is buffer-based, where data is downloaded in bursts depending on the streaming algorithm and playback state.

\begin{figure}[htbp]
    \centering
    \begin{subfigure}[b]{\linewidth}
        \centering
        \includegraphics[width=2in, height=1.5in]{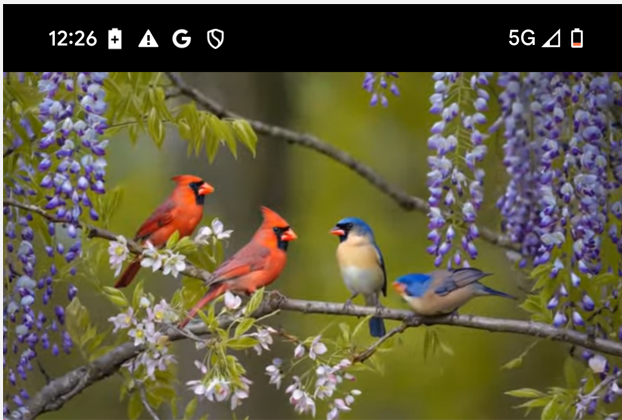}
        \caption{Single UE}
        \label{youtube1}
    \end{subfigure}
    
    \vspace{0.5cm}
    
    \begin{subfigure}[b]{\linewidth}
        \centering
        \includegraphics[width=3.5in, height=1.5in]{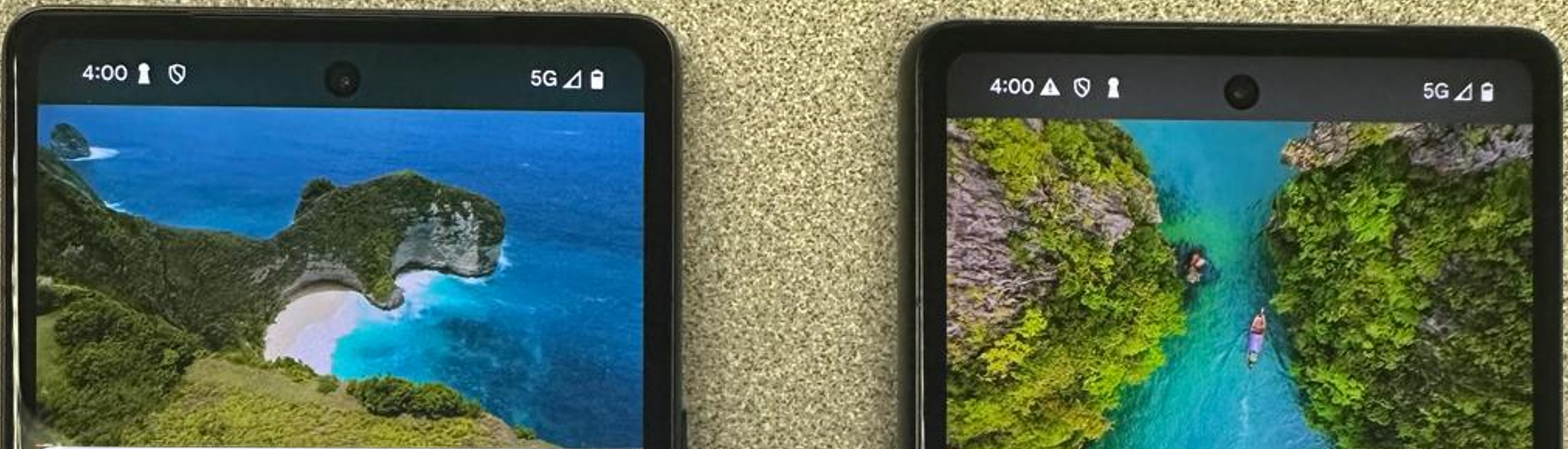}
        \caption{Multiple UEs}
        \label{youtube2}
    \end{subfigure}
    \caption{YouTube-based data generation}
    \label{youtube}
\end{figure}

Each sample in the dataset captured a snapshot of the underlying radio conditions and MAC scheduler behavior. Standard preprocessing was applied, including timestamp alignment, outlier removal, and feature normalization, to prepare the dataset for supervised regression modeling.

\section{Result and Analysis} 
\begin{figure}[htbp]
    \centering
    \begin{subfigure}[b]{\linewidth}
        \centering
        \includegraphics[width=0.8\linewidth,height=5cm]{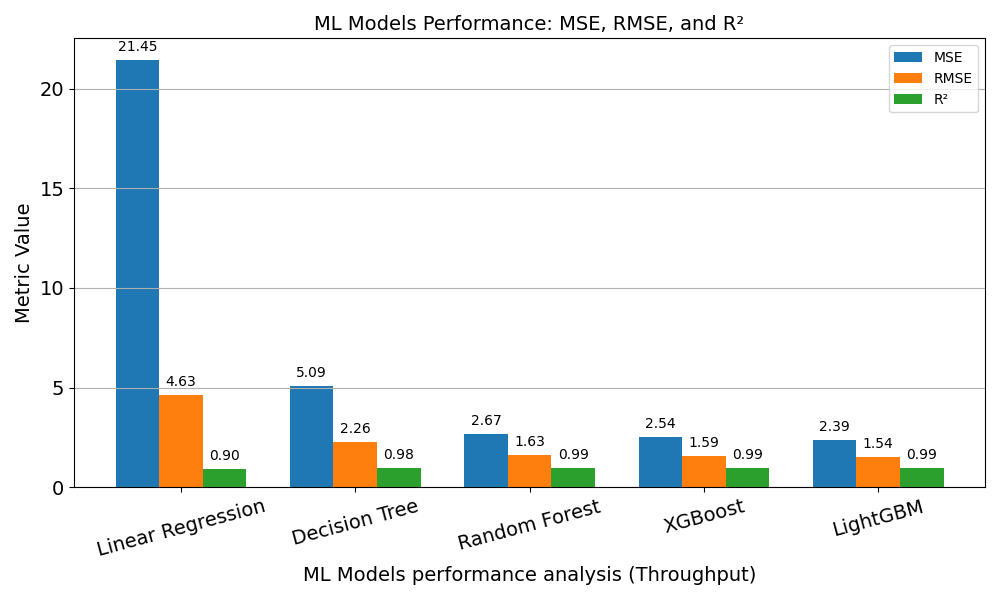}
        \caption{Throughput performance analysis}
        \label{model_performance1}
    \end{subfigure}
    
    \vspace{0.5cm}
    
    \begin{subfigure}[b]{\linewidth}
        \centering
        \includegraphics[width=0.8\linewidth, height=5cm]{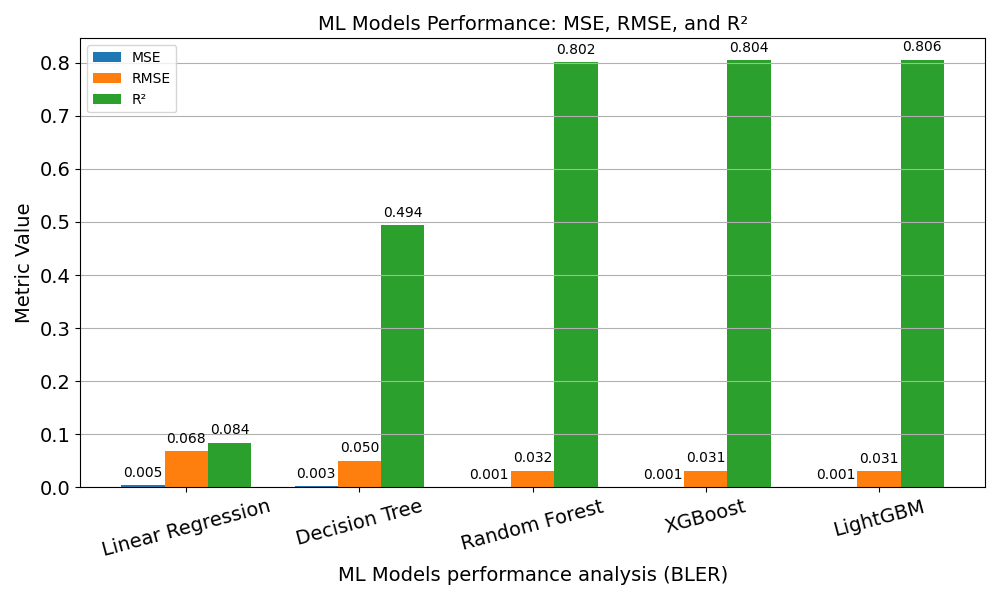}
        \caption{BLER performance analysis}
        \label{model_performance2}
    \end{subfigure}
    \caption{ML models performance analysis}
    \label{ml_models}
\end{figure}

The performance of the five machine learning models — Linear regression, Decision tree regression, Random forest regression, XGBoost, and LGBM — was evaluated using the 5G SA test dataset. The models were compared using MSE, RMSE, and $R^2$ score. The results are presented in Table \ref{table1}, and a grouping bar chart visualization is shown in Fig. \ref{ml_models} for clarity.

\begin{table}[htbp]
    \centering
    \caption{ML Models Comparison}
    \begin{adjustbox}{width=0.45\textwidth}
    \begin{tabular}{|l|c|c|c|}
        \hline
        \textbf{Model} & \textbf{MSE ($kbps^2$)} & \textbf{RMSE (kbps)} & \textbf{$R^2$ Score}\\
        \hline
        Linear Regression & 21.45 & 4.63 & 0.90 \\
        \hline
        Decision Tree & 5.09 & 2.26 & 0.98 \\
        \hline
        Random Forest & 2.67 & 1.63 & 0.99\\
        \hline
        XGBoost & 2.54 & 1.59 & 0.99\\
        \hline
        LightGBM & 2.39 & 1.54 & 0.99\\
        \hline
    \end{tabular}
    \end{adjustbox}
    \label{table1}
\end{table}

Figs. \ref{model_performance1} and \ref{model_performance2} compare the performance of five regression models—Linear Regression, Decision Tree, Random Forest, XGBoost, and LGBM —evaluated using MSE, RMSE, and $R^2$. The two performance analyses correspond to cell throughput and link BLER performance metrics. As shown in Fig. \ref{model_performance1}, which is a throughput analysis, all models achieve lower MSE and RMSE values, along with significantly higher $R^2$ scores. Notably, the LGBM model demonstrates the best overall performance with the lowest MSE and RMSE metrics (MSE $\approx$ 2.39, RMSE $\approx$ 1.54) and the highest $R^2$ ($\approx$ 0.99), indicating a strong correlation between predicted and actual throughput. Fig. \ref{model_performance2}, which analyzes the BLER performance metric, also demonstrates LGBM performs best among these ML models. MSE and RMSE values are higher, and $R^2$ scores are lower than LGBM regression, indicating poor models fit for rest of the ML models. 

\begin{figure}[htbp]
    \centering
    \begin{subfigure}[b]{\linewidth}
        \centering
        \includegraphics[width=0.8\linewidth,height=5cm]{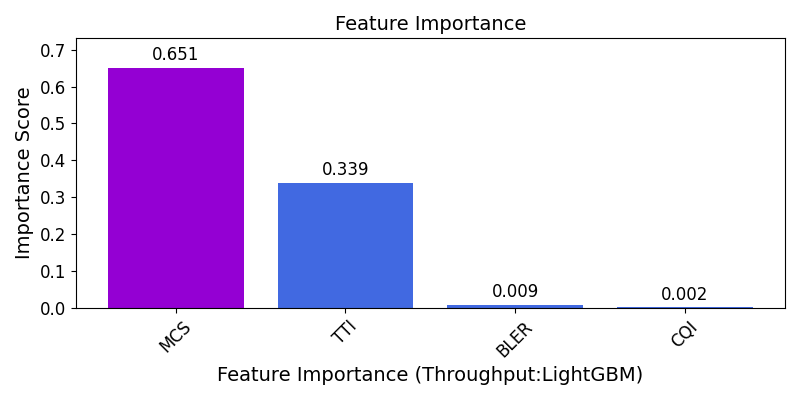}
        \caption{Feature importance (Throughput)}
        \label{fig:throughput}
    \end{subfigure}
    
    \vspace{0.5cm}
    
    \begin{subfigure}[b]{\linewidth}
        \centering
        \includegraphics[width=0.8\linewidth, height=5cm]{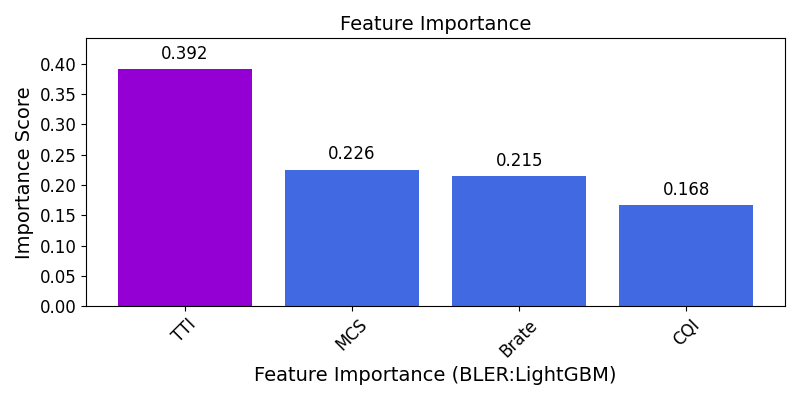}
        \caption{Feature importance (BLER)}
        \label{fig:error}
    \end{subfigure}
    \caption{Feature importance analysis}
    \label{features}
\end{figure}
To better understand which features most influence throughput, we extracted feature importance scores from the LGBM model. As shown in Fig. \ref{features}, MCS was the most influential parameter, followed by TTI, BLER, and CQI. This indicates that throughput in 5G SA networks is influenced not only by channel quality but also significantly by how much MCS and TTI are allocated — emphasizing the need for MCS and TTI-aware optimization strategies. However, for BLER performance analysis, we see TTI has a strong influence as compared to other feature metrics such as MCS, Brate, and CQI.

\begin{figure}[htbp]
    \centering
    \begin{subfigure}[b]{\linewidth}
        \centering
        \includegraphics[width=0.8\linewidth,height=5cm]{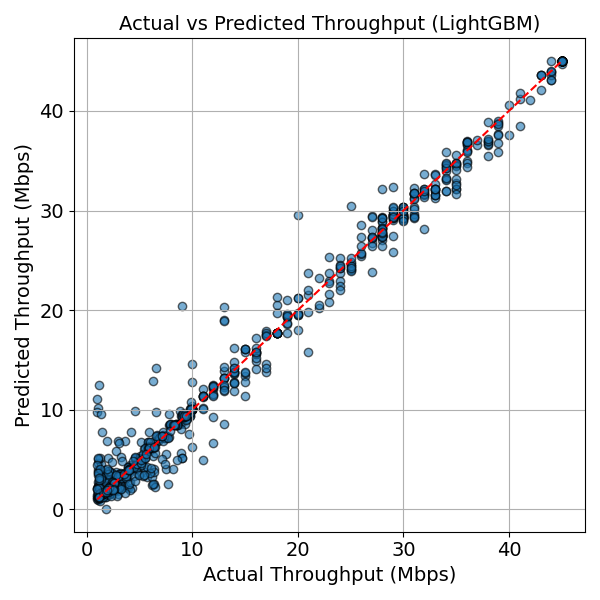}
        \caption{Actual vs predicted (throughput)}
        \label{actual_vs_predicted1}
    \end{subfigure}
    
    \vspace{0.5cm}
    
    \begin{subfigure}[b]{\linewidth}
        \centering
        \includegraphics[width=0.8\linewidth, height=5cm]{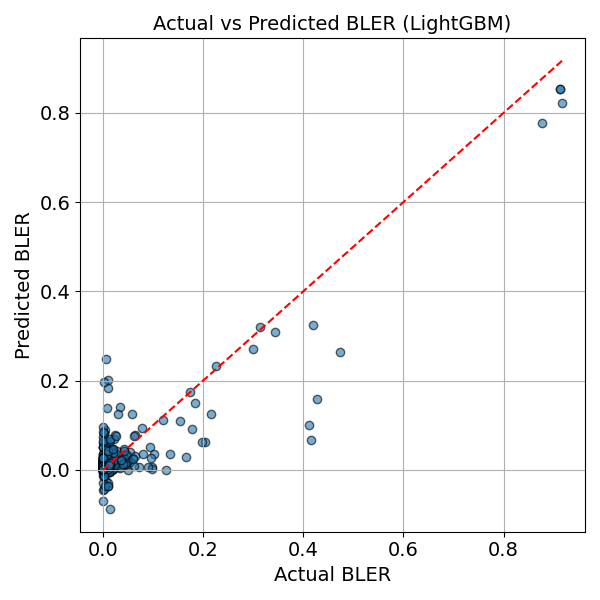}
        \caption{Actual vs predicted (BLER)}
        \label{actual_vs_predicted2}
    \end{subfigure}
    \caption{Actual vs predicted throughput and BLER analysis}
    \label{actual_vs_predicted}
\end{figure}

Figs. \ref{actual_vs_predicted1} and \ref{actual_vs_predicted2} illustrate the relationship between actual and predicted throughput and BLER values using an LGBM model. The scatter plots represent the performance of the model under two configurations: for throughput performance analysis (Fig. \ref{actual_vs_predicted1}) and for BLER performance analysis (Fig. \ref{actual_vs_predicted2}). The red dashed line in both plots denotes the ideal diagonal where predicted values match actual measurements perfectly. In Fig. \ref{actual_vs_predicted1}, where MCS, TTI, BLER, and CQI are included as input features, the predicted throughput values exhibit tighter clustering around the ideal diagonal, throughout the 1–40 Mbps range. This indicates improved model generalization and higher prediction accuracy. The predictions more accurately reflect actual throughput even in cases of variation across different link conditions, suggesting that MCS is a critical feature for capturing throughput dynamics. In contrast, Fig. \ref{actual_vs_predicted2}, which includes MCS, TTI, Brate, and CQI as feature vectors, shows a broader dispersion of points and a tendency to over- or under-predict BLER at various operating points. The prediction accuracy degrades especially at mid BLER values, indicating the model's limited ability to capture BLER-related variance. This performance gap highlights the importance of including adaptive bit rate as a feature in machine learning models for BLER prediction.

\begin{figure}[htbp]
    \centering
    \begin{subfigure}[b]{\linewidth}
        \centering
        \includegraphics[width=0.8\linewidth,height=5cm]{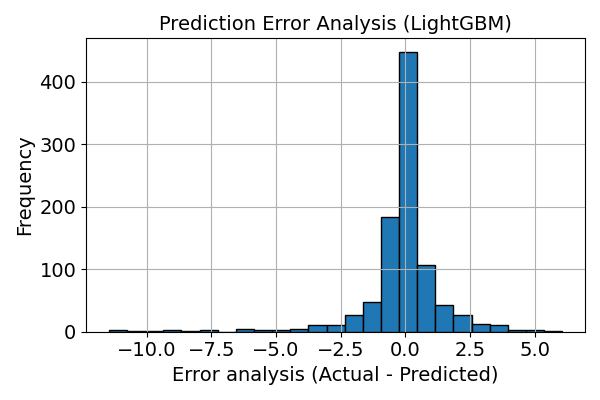}
        \caption{Prediction error analysis (throughput)}
        \label{prediction_error_distribution1}
    \end{subfigure}
    
    \vspace{0.5cm}
    
    \begin{subfigure}[b]{\linewidth}
        \centering
        \includegraphics[width=0.8\linewidth, height=5cm]{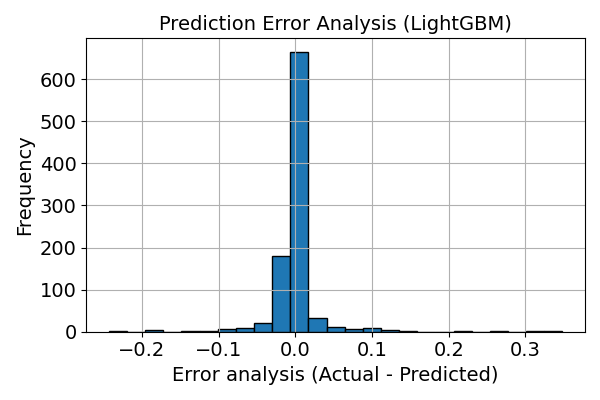}
        \caption{Prediction error analysis (BLER)}
        \label{prediction_error_distribution2}
    \end{subfigure}
    \caption{Prediction error analysis}
    \label{prediction_error_distribution}
\end{figure}

Figs. \ref{prediction_error_distribution1} and \ref{prediction_error_distribution2} present the prediction error distributions of the LGBM model under two configurations: throughput analysis and BLER analysis. The prediction error is defined as the difference between the actual and predicted throughput (BLER) values for each test instance. As shown in Fig. \ref{prediction_error_distribution1}, when throughput analysis is incorporated, the prediction error distribution is sharply centered around zero, with the majority of errors falling within a narrow range of ±2.5 Mbps. This indicates low variance and high confidence in the model’s predictions, confirming the importance of TTI and MCS as the discriminative features for throughput estimation. The distribution also shows minimal skewness or heavy tails, which suggests that the model generalizes well across the dataset. In contrast, Fig. \ref{prediction_error_distribution2}, which deals with BLER analysis, reveals a shorter and narrower error distribution, with most prediction errors within ±0.05 Mbps. As most of the samples are within very narrow range, our BLER analysis performs even better than the throughput analysis.

Together, these results reinforce that incorporating TTI adaptation into the feature space not only improves average predictive accuracy, but also significantly reduces the dispersion and extremity of prediction errors. This has practical implications for robust link quality estimation and proactive resource management in adaptive wireless systems.

\section{Conclusion}
This paper presents a machine learning-based framework for predicting downlink throughput and BLER in 5G NR networks using real-world measurements collected from two smartphone UEs connected to an srsRAN-based testbed. The dataset included physical layer indicators (CQI, MCS, TTI, Bit rate, and BLER), captured under diverse channel conditions, including LOS, nLOS, and mobility scenarios. Five regression models — Linear Regression, Decision Tree, Random Forest, XGBoost, LGBM — were evaluated using MSE, RMSE, and $R^2$ metrics. Among these, LGBM consistently outperformed the others both throughput and BLER analysis, confirming its ability to model complex, nonlinear dependencies in realistic 5G environments. Feature importance analysis further revealed that MCS and TTI are the most influential factors, underscoring the importance of resource allocation awareness in prediction analysis. The proposed framework enables throughput and BLER estimation using only observable radio features, offering a practical foundation for real-time quality-of-service (QoS) prediction and adaptive resource management. Our work includes joint throughput and BLER prediction, model generalization across different scenarios, and its the foundation block for future research in 5G NR and beyond.

\bibliography{oran}

\end{document}